\documentclass{llncs}
\usepackage[T1]{fontenc}
\usepackage{graphicx}
\usepackage{hyperref}
\usepackage{tabu}                      
\usepackage{booktabs}                  
\usepackage[table,x`cdraw]{xcolor}

\usepackage{pifont}
\newcommand{\RVone}[1]{ {\color{black!100} #1}}
\begin{document}
\title{A Literature Review and Taxonomy of In-VR Questionnaire User Interfaces}
%
%
\author{Saeed Safikhani \inst{1} \and
Lennart Nacke \inst{2} \and
Johanna Pirker \inst{1,3}}

\institute{Graz University of Technology, Graz, Austria \and
University of Waterloo, Waterloo, Canada \and 
Ludwig Maximilian University, Munich, Germany}

\authorrunning{S. Safikhani et al.}
%
%
\maketitle              
\begin{abstract}
Previous research demonstrates that the interruption of immersive experiences may lead to a bias in the results of questionnaires. Thus, the traditional way of presenting questionnaires, paper-based or web-based, may not be compatible with evaluating VR experiences.
Recent research has shown the positive impact of embedding questionnaires contextually into the virtual environment. However, a comprehensive overview of the available VR questionnaire solutions is currently missing. Furthermore, no clear taxonomy exists for these different solutions in the literature.
To address this, we present a literature review of VR questionnaire user interfaces (UI) following PRISMA guidelines. Our search returned 1.109 initial results, which were screened for eligibility, resulting in a corpus of 25 papers. This paper contributes to HCI and games research with a literature review of embedded questionnaires in VR, discussing the advantages and disadvantages and introducing a taxonomy of in-VR questionnaire UIs.

\end{abstract}
\section{Introduction}
Virtual reality (VR) remained a specialized technology for much of the last century until---in the previous decade---lighter-weight, consumer-friendly head-mounted displays (HMDs) were introduced, and games were pulling early adopters and industries into this space \cite{nascimento2019role,goedicke2018vr,safikhani2022BIM,safikhani2021virtual,pottle2019virtual,sidani2021recent,di2020review,barnard_2022}.
Many researchers use VR to conduct user studies using either objective methods and sensory data (e.g., respiration sensor, brain activity measurement, blood pressure, eye tracking \cite{putze2020breaking,Souvik2020,hertweck2019brain}), or subjective methods and questionnaires (e.g., PXI~\cite{Abeele2020}, IPQ \cite{schubert2001experience}, PENS \cite{ryan2006motivational}).
However, the difficulties in applying objective methods lead to the increasing use of subjective methods, especially questionnaires, for the evaluation process. \cite{alexandrovsky2020examining,bruehlmann2018surveys,harris2019virtually}.

The traditional methods for completing questionnaires have been either paper-based forms or web-based user interfaces. Although these ways of presenting questionnaires have been used for a long time and are well-developed for user needs, the immersive features of VR may not be compatible with them. 
Recent studies show that when the subject of the questionnaire is related to a feature of experience that can be interrupted by exiting the environment (e.g., perception of presence), implementing the questionnaire inside the environment can lead to better consistency and reliability of the results \cite{regal2019questionnaires,alexandrovsky2020examining,putze2020breaking} as well as an increase in motivation, engagement, and presence of participants \cite{steinmaurer2021engagement,frommel2015integrated}.

These influences may be even more dominant for immersive VR experiences, as one of the main features is the high degree of immersion \cite{safikhani2021influence}.
For example, imagine a situation where the research objective is to measure presence during a psychiatric therapy session in VR. 
The traditional methods would require participants to leave the virtual environment and remove the headset to complete the questionnaires. This procedure can lead to an interruption of the user's perception of presence and biased answers.
Considering these points, recent studies investigate the influence of embedding questionnaires as an element directly in the virtual environment. This element can be presented as a separate user interface (UI) or as an interactive object in VR \cite{safikhani2021influence}.
Accordingly, it is crucial to understand alternative methods of presenting questionnaires for VR studies to overcome the barriers of traditional workflows. 

Evaluating the impact of embedding questionnaires in a VR environment can improve the research communities' understanding of its benefits and limitations. Furthermore, enhancing the implementation procedure and removing barriers limiting the use of embedded questionnaires may lead to increased uptake.
Daily increasing tendency to use VR applications and metaverse daily \cite{Youn2021Metaverse}, emphasis a need for more detailed studies on VR. On the other hand, the traditional way of using questionnaires in VR may lead to biased results. 
In this paper, we review recent articles that either use embedded questionnaires as a method for their studies or as a research topic to investigate and improve this type of questionnaire representation and to answer the following questions:

\begin{itemize}
    \item{\textbf{RQ1:}} How does the literature refer to or define in-VR questionnaires?
    \item{\textbf{RQ2:}} What are the most commonly used UIs and interactions for in-VR questionnaires?
    \item{\textbf{RQ3:}} What are future research and development opportunities for in-VR questionnaires?
\end{itemize}
We use the term ``in-VR,'' referring to questionnaires embedded and integrated into VR. The term ``out-VR'' refers to the traditional paper-based or web-based representation of questionnaires.
While our literature review points to works exploring different aspects of in-VR questionnaires, a precise overview of how they can be presented in virtual environments and how to interact with them is lacking. Furthermore, the terminology used to refer to in-VR questionnaires is imprecise.
We contribute to human-computer interaction (HCI) research in games with a taxonomy of in-VR questionnaires, discussing common ways to present and interact with them in VR. It is important to investigate the impact of various design choices on the user experience of virtual reality (VR) in education. This is because the use of VR technology in education is rapidly increasing. To achieve this, using questionnaires to gather user feedback will be necessary. This paper can be useful in future studies on VR educational content by providing a review of potential VR questionnaire terminology and designs. 

\section{Related Work}

\RVone{A taxonomy is a collection of dimensions, each composed of features \cite{hertel2021taxonomy}. Examining related work on the taxonomy of various attributes of virtual reality experiences can help researchers in the HCI community identify current research frontiers and future directions. A review by Kim et al. \cite{kim2020systematic} on 65 articles, categorizes the affecting factors on the user experience (UX) in VR based on user characteristics, device settings, and interaction types. In addition to these factors, the evaluation method can also influence the results and their interpretation. 

Motejlek and Alpay \cite{motejlek2021taxonomy} did a taxonomy focused on the application of virtual and augmented reality in education and based on FURPS+ \cite{eeles2005capturing} method. Their study results in categorizing the VR experience based on user interactions in four classes: General Purpose Controller, User Tracking, Specialized Controller, and No Interaction.
Most VR interactions can be performed with gestures. In a study by Bhowmick et al. \cite{bhowmick2021understanding}, a taxonomy of gestures for object selection in VR was developed. This taxonomy can be useful for interacting with different UI elements. According to their taxonomy, gestures in VR can be categorized into two main types: Hand dominance and movement with multiple body parts. Multiple body part movement can be as simple as walking or combining multiple body part gestures (this may require additional sensors to track body parts). Hand dominance gestures can rely only on the dominant hand or be performed in combination with gestures from the non-dominant hand (developers may also not consider a dominant hand).
Another taxonomy by Ruscella and Obeid \cite{ruscella2021taxonomy} tries to categorize the attributes of VR applications that lead to a successful immersive experience. Accordingly, they categorize the influencing factors in: Interactivity, Story, Gamification, Dynamics, Co-participation, Embodiment, Immersive Technology, Meta control, and Didactic Capacity.}
To analyze these factors, collecting and analyzing feedback from participants in user research studies can be an essential step\cite{lazar2017research}.
Most HCI studies in recent years have addressed empirical and artifact contributions \cite{wobbrock2016research,shneiderman2011claiming}.

If the research topic is to consider the measurement of values related to participants' beliefs and feelings about particular topics, researchers may use self-report measures (i.e., questionnaires) \cite{field2002design}.
A questionnaire may contain standardized and structured questions based on previous studies to elicit specific results or free-form questions to reflect user preferences without limitation of standard questions \cite{rosenthal2008essentials}.
As with other research methods, the influencing factors that lead to bias in the results should be reduced in the design and implementation of the questionnaire. 
In a study by Choi et al. \cite{choi2005peer}, 48 types of bias in self-report surveys are discussed and divided into three main types: Question Design, Questionnaire Design, and Questionnaire Administration. 
In the case of immersive experiences, Alexandrovsky et al. \cite{alexandrovsky2020examining} argue, based on a study by Schwind et al. \cite{schwind2019using}, that consistency in questionnaire administration can lead to a reduction in random errors.
An example of these errors can be conditions that lead to a disruption of the sense of presence by switching between real and virtual environments.

A study by \cite{knibbe2018dream} shows that leaving a virtual environment can lead to disorientation. The extent of this disorientation depends on the duration of the VR experience, the degree of presence in the VR environment, and the distance traveled. They found that this disorientation even occurs in simplified virtual environments.
These types of interruptions can negatively impact user engross involvement (a level of involvement in the virtual environment that emotionally affects the user through the experience) and lead to less reliable measurements \cite{alexandrovsky2020examining}.
Thus, if we can keep the user in VR during the subjective measurement period of the study, we could reduce the negative effects of leaving the virtual environment.
Here is the point that concept of in-VR questionnaires can be defined to provide the possibility of doing questionnaires inside a VR application without leaving the immersive experience to reduce these errors.

In recent years, attention to in-VR questionnaires has increased, and several researchers have used them in their studies, e.g., \cite{alexandrovsky2020examining,kang2018flotation,cao2018visually,schwind2018up,schwind2019using,fernandes2016combating,wienrich2018social}.
\cite{schwind2019using} reports that there is no significant difference between the results of the in-VR and out-VR questionnaire in the case of presence. A similar result is also reported by \cite{alexandrovsky2020examining}, which confirms the previous results. Although \cite{schwind2019using} showed that embedding the questionnaire in the VR can improve the consistency of the results compared to out-VR, \cite{alexandrovsky2020examining} found no significant difference. This inconsistency in the results leaves the discussion of the influence of the in-VR questionnaire design on the reported results open for further investigation.

In addition, there is no clear guide for implementing in-VR questionnaires considering UI and interaction systems.
Although 3D UI design has great potential to use in VR applications \cite{laviola20173d}, most implemented ones replicate the UI of traditional (non-immersive) desktop experiences. On the other hand, the consistency of the interaction method with the type of content in VR can lead to a better user experience \cite{HEPPERLE2019321}. It may also reduce the time needed to interact with the UI, which can be important for long questionnaires in VR.
Alexandrovsky et al. \cite{alexandrovsky2020examining} reviewed 123 articles from 2016 to 2019 that addressed the use of questionnaires in VR studies and, in particular, in-VR questionnaires. They identified 15 articles that used in-VR questionnaires. These embedded questionnaires used different presentation methods such as heads-up display, world-based UI, and body-based UI.
Their results show diversity in implementing in-VR questionnaires and a lack of comprehensive study.

In this paper, we aim to develop a taxonomy for the design of in-VR questionnaires.
In recent years, there are articles that address the review or implementation of in-VR questionnaires, such as the study by Alexandrovsky et al. \cite{alexandrovsky2020examining}. While these studies use literature review and/or implementation of in-VR questionnaires, they do not introduce a clear term for these UI elements. The lack of a general term may lead to less use of In-VR questionnaires. In this study, we propose a general term for the user interface of these questionnaires after reviewing the collected literature. Moreover, the study by Alexandrovsky et al. \cite{alexandrovsky2020examining} only considered the traditional user interface for In-VR questionnaires (the common user interface adopted from desktop applications). However, in this review, we expand the UI design options by considering intradiegetic and 3D UI elements, e.g. \cite{safikhani2021influence,wagener2020investigating}.
Accordingly, we identified two dimensions for this taxonomy: UI and interaction type. We reviewed the screened articles to find a relevant application of these two dimensions to use embedded questionnaires in VR. This taxonomy can help the HCI community to choose an appropriate UI and interaction when designing an in-VR questionnaire based on their application. 

\section{Methodology}
We conduct a literature review using the Preferred Reporting Items for Systematic Reviews and Meta-Analyses (PRISMA) guidelines \cite{page2021prisma} for the paper screening procedure.
We extracted the literature for this review using \href{https://www.scopus.com}{Scopus} (one of the largest abstract and citation databases of peer-reviewed literature that can search in various libraries such as ACM, Springer, and ScienceDirect) and \href{https://origin-ieeexplore.ieee.org}{IEEEXplore} (as a common database for computer science and technology-related articles). Our search terms in both databases are: \texttt{(``in-VR'' OR ``in virtual reality'') AND (``Questionnaire'' OR ``Survey'')}.
The first author of this paper collected the literature from the defined databases.
We consider a data range for this literature gathering from 01.01.2014 to 31.12.2021 that leads to 1109 records (976 from Scopus and 133 from IEEEXplore).
We can summarize our exclusion criteria as follows:(1) The papers that were published before 2014, (2) did not have an appropriate format (included only an abstract or preview), (3) duplicates, (4)papers that were written in languages other than English, (5) papers that were described experiences without HMDs, (6) papers to which we did not have access
We included articles that had ``in VR questionnaire'' or ``inVRQ'' (common terms for embedded questionnaires in VR) in their title or keywords, then searched for them in the abstract, and finally searched for related content to in-VR questionnaire in the full text.
We performed the screening process manually and did not use any automation tool.
We performed an informal quality assessment of the contribution content focused on the application of the in-VR questionnaire and the implementation methods. As reviewing this topic is rather new, for this topic, we utilized inductive thematic analysis to explore new insights and perspectives on its taxonomy and application.
Accordingly, from 734 articles published between 2014 and 2021 (after the release of consumer-based VR device to the time of data collection), we selected 25 articles for further discussion in this paper that considered either the in-VR questionnaire as an assessment tool or as a topic for their study. Among these articles, nine reported on user studies using in-VR questionnaires, one developed only a toolkit without a user study, and two were review papers.
In \autoref{tab:literature}, we summarize the reviewed articles in the literature considering their topic, the number of participants, and the implemented questionnaires.

\begin{table}[htb]
\begin{tabular}{p{0.1\linewidth}  p{0.45\linewidth}  p{0.01\linewidth}  p{0.25\linewidth}}
\toprule
\multicolumn{1}{c}{Paper} & \multicolumn{1}{c}{Type of study}                  & \multicolumn{1}{c}{Participants} & \multicolumn{1}{c}{Questionnaires}      \\ \midrule
\cite{alexandrovsky2020examining}       & Literature review, study on UI and interaction     & 10                                         & IPQ, NASA TLX, UMUX, SUS                                                       \\
\cite{bushra2018comparative}           & Study with application of inVRQ         & 13                                        & SSQ and SUS                                                                \\
\cite{cao2018visually}           & Study with application of inVRQ         & 22                                        & SSQ                                                                \\
\cite{cha2019spatial}           & Study with application of inVRQ         & 43                                        & Custom                                                                \\
\cite{degraen2019enhancing}           & Study with application of inVRQ          & 10                                        & Custom   \\
\cite{feick2020virtual}                & Toolkit Development                                & -                                          & NASA TLX, SUS, SSQ                                                                          \\
\cite{fernandes2016combating}           & Study with application of inVRQ         & 32                                       & Discomfort score                                                                \\
\cite{graf2020inconsistencies}               & Study on presence in VR                            & 36                                         & IPQ, SUS                                                                                   \\
\cite{kang2018flotation}           & Study with application of inVRQ         & 13                                        & SSQ and SUS                                                                \\
\cite{kopel2021gameplay}               & Study on immersion in VR                           & 26                                         & Custom                                        \\
\cite{krekhov2019deadeye}            & User study with application of inVRQ & 24                                         & NASA TLX, custom                          \\
\cite{luong2021survey}                & Study with application of inVRQ    & -                                          & -                                    \\
\cite{marin2018affective}           & Study with application of inVRQ         & 60                                       & SAM                                                                \\
\cite{oberdorfer2019usability}           & Study with application of inVRQ         & 13                                        & NASA TLX                                                                 \\
\cite{putze2020breaking}               & Study on the influence of inVRQ      & 36                                         & IPQ, SUS, PQ                                                                           \\
\cite{regal2018vrate}           & Toolkit Development          & -                                      & -                                                                \\
\cite{safikhani2021influence}           & Study on inVRQ UI design             & 16                                         & NASA TLX, IPQ, SUS                                                                 \\
\cite{schwind2019using}           & Study on the influence of inVRQ         & 36                                        & SUS, WS, IPQ                                                                 \\
\cite{schwind2017these}           & Study with application of inVRQ         & 28                                       & WS and custom                                                                \\
\cite{tamaki2021shoot}             & Study on the influence of inVRQ      & 12                                         & IPQ                                                                                     \\
\cite{vieira2019knowledge}           & Study with application of inVRQ         & 35                                        & Custom                                                                 \\
\cite{wagener2020investigating}            & Study on inVRQ UI design             & 35                                         & Presence and custom  \\
\cite{wienrich2018social}           & Study with application of inVRQ         & 12                                       & PANAS and presence                                                               \\
\cite{wijnen2020performing}               & User study with application of inVRQ & 38                                         & Custom                           \\
\cite{wolfel2021entering}              & Literature review                                  & -                                          & -                                                        \\
\bottomrule
\end{tabular}
\caption{ An overview of the collected literature, their questionnaires and participants}
\label{tab:literature}
\end{table}



\section{Results}


Our review indicates, most articles considered mouse pointer interaction as one of the typical interaction types in VR. They used a traditional 2D user interface (similar to the user interface in out-VR) for in-VR questionnaires.
Most articles used standard or custom questionnaires with multiple-choice responses, such as a 5-point Likert scale. We have organized the literature we gathered into three main themes: questionnaires, design, and toolkit. Articles that focused on questionnaires mainly discussed the use of in-VR questionnaires and compared them to out-VR questionnaires. Papers that focused on design discussed various considerations for designing in-VR questionnaires. Finally, some papers provided or recommended a toolkit for implementing in-VR questionnaires.

\textbf{Questionnaires:}
One of the main characteristics of VR is a high level of immersion that leads to a sense of presence.
According to Slater, \cite{slater2003note} \textbf{immersion} can be defined as the objective perception of a virtual experience that is influenced by device technology. Higher sensory displays and tracking technology that lead to a more consistent experience with real sensory perception can lead to higher immersion.
On the other hand, presence is the subjective feeling of "being" in a virtual environment. Given the same level of immersion, we can expect different degrees of presence depending on the user's response.
Therefore, measuring the sense of presence is crucial for VR research and development. Completing the questionnaire outside the virtual environment may lead to disruption of the feeling of presence because of time consumption and disorientation, it is essential to compare the effect of questionnaires in VR and outside VR on user experience and outcomes. 

Schwind et al. \cite{schwind2019using} examined the effects of embedding questionnaires in a virtual environment on participant response. A total of 36 participants took part in the study by playing a first-person shooter game in VR.
They found that when questionnaires are used in the virtual world, the variance of the measurement can be kept constant, but this is not the case when questionnaires are used outside the virtual world. This difference could be due to the fact that the probability of a ``Break In Presence'' (BIP) is lower when using in-VR questionnaires. 
BIP can be defined as a moment when the sense of presence is disturbed and the user becomes aware of the real environment \cite{chung2010analysis}. A sudden change of context or taking off the headset to answer the questionnaire can lead to BIP, which can bias the assessment results \cite{putze2020breaking}. The effects of BIPs on performance and presence may vary depending on the complexity and workload of the VR tasks, as well as the type of VR environment (e.g., learning, productivity, therapy). Embedding questionnaires directly into the VR environment can contribute to the consistency of self-reports on presence, but the extent and influence of BIPs on the VR experience need to be further investigated. A study by Tamaki and Nakajima \cite{tamaki2021shoot} shows that some items of the IPQ (e.g., ``I am in the VR space'' and ``I am involved in the virtual environment'') decrease when users experience the transition between VR and out-VR questionnaires. Putze et al. \cite{putze2020breaking} discussed the effects of in-VR questionnaires on BIPs compared to out-VR questionnaires. They recorded biosignals from 50 participants in VR games with two environmental designs, one realistic and one simplified. The results of this study show that both in-VR and out-VR questionnaires can cause BIPs. However, the in-VR questionnaires produce less intense BIPs and may result in more reliable responses.

Although questionnaires are well-known tools for assessing presence in virtual environments, a recent study by Graf and Schwind \cite{graf2020inconsistencies} shows inconsistencies between measurement and main results. This finding is in agreement with Alexandrovsky et al. \cite{alexandrovsky2020examining} findings that presence questionnaires are not sufficient to measure BIPs. They recommend measuring presence in the virtual environment, either behaviorally or physiologically. The study by Graf and Schwind \cite{graf2020inconsistencies} also shows that some results are influenced by the environment in which the questionnaires are placed and the type of interaction with them.
Accordingly, they recommend a revision of the presence assessment tool in VR and show the importance of studying the influence of the design of the questionnaire environment on users' responses.
However, this result contradicts the results of Regal et al. \cite{regal2019questionnaires} study on the influence of using a special environment for questionnaires or using questionnaires as a scene object. Regal et al. \cite{regal2019questionnaires} report that they could not find any influence of the questionnaire environment on the result of the presence questionnaire. 

\textbf{Design:}
Using questionnaires in VR can be a research topic in itself focused on the design and development of these virtual elements. Here, the lack of clear guidelines and standards for 3D UIs leads to many different UIs for questionnaires in the literature. Alexandrovsky et al. \cite{alexandrovsky2020examining} conducted user studies to evaluate different design choices for in-VR questionnaires. They began the study by surveying VR experts about their attitudes toward VR questionnaires. The general response was positive, but there was no standard method for presenting questionnaires in the virtual environment. 

Moreover, in contrast to the positive tendency of the experts to embed questionnaires in VR, there were also some arguments against it. These arguments are mainly related to the difficulties in designing/implementing questionnaires in VR and overloading the participants during the survey, which may lead to the bias of the results.
The authors then discussed a conducted user study in which they designed two positions (world-anchored and body-anchored) for questionnaire placement and implemented two interaction types (pointer and trackpad). Their results show that placing the questionnaire in a fixed position in the virtual environment with pointer interaction works better than the other options. 
This result is consistent with the results of Safikhani et al. \cite{safikhani2021influence}, where the authors compared two alternatives for designing questionnaires in VR with those outside of VR. They considered two world-anchored objects for displaying questionnaires in the virtual environment and called them 2D and 3D layouts. The first alternative, the 2D layout, is a monitor on the wall that displays the question text and allows the user to select the answer with a pointer. In the second case, the 3D layout, they designed a virtual device with multiple monitors and handles to display the text and answer choices. Users can select, accept, and change the answer using the handles via VR grip interaction. Their results suggest that most participants prefer the in-VR design of questionnaires. However, task load is higher compared to out-VR, which is consistent with \cite{alexandrovsky2020examining} findings.

Even when questionnaires are embedded in the virtual environment, the consistency of the design with the environment can affect the user experience. Wagener et al. \cite{wagener2020investigating} investigated the impact of the extent of integration and consistency of the questionnaire with the virtual environment on the user experience and the duration of the entire study. They considered four design alternatives for the questionnaire UI and categorized them as extradiegetic and intradiegetic. Their results suggest that an intradiegetic design approach can lead to higher presence, perceived user experience, and expert acceptance. However, the time required to complete the questionnaires is higher. A similar finding is also reported by Safikhani et al. \cite{safikhani2021influence} that questionnaires with 3D layouts are more time-consuming.

\textbf{Toolkit:}
Considering the additional effort required to implement questionnaires in the virtual environment, researchers have developed toolkits to facilitate this process. Feick et al. \cite{feick2020virtual} has developed a customizable toolkit to quickly implement a basic user interface, similar to the traditional desktop user interface, in VR. Researchers can use either a set of standard predefined questionnaires, such as NASA TLX and SUS, or custom questionnaires with JSON files. Safikhani et al. \cite{safikhani2021influence} accompanied their research with an immersive questionnaire toolkit. With their toolkit, researchers can implement standard or custom questionnaires, either based on XML files or by importing from the Limesurvey website (\url{limesurvey.org}). With this toolkit, evaluation results can later be exported in XML file format or uploaded to the Limesurvey website for post-processing.
We found also articles using in-VR questionnaires as an evaluation tool for their studies. For example, Wijnen et al. \cite{wijnen2020performing} simulate a virtual museum tour where participants encounter interacting robots. They used a custom in-VR questionnaire to assess the feeling of interacting with robots in VR and the perceived scene of presence. In this study, the in-VR questionnaire was presented as a 2D UI object fixed at a specific location in the scene. Participants can interact with the questionnaire through pointer interaction. The same type of interaction is used by Krekhov et al. \cite{krekhov2019deadeye}, but a full-screen 2D UI was used to present NASA TLX and a custom questionnaire. 
\subsection{Classification} 
\begin{figure*}
    \centering
    \includegraphics[width=\textwidth]{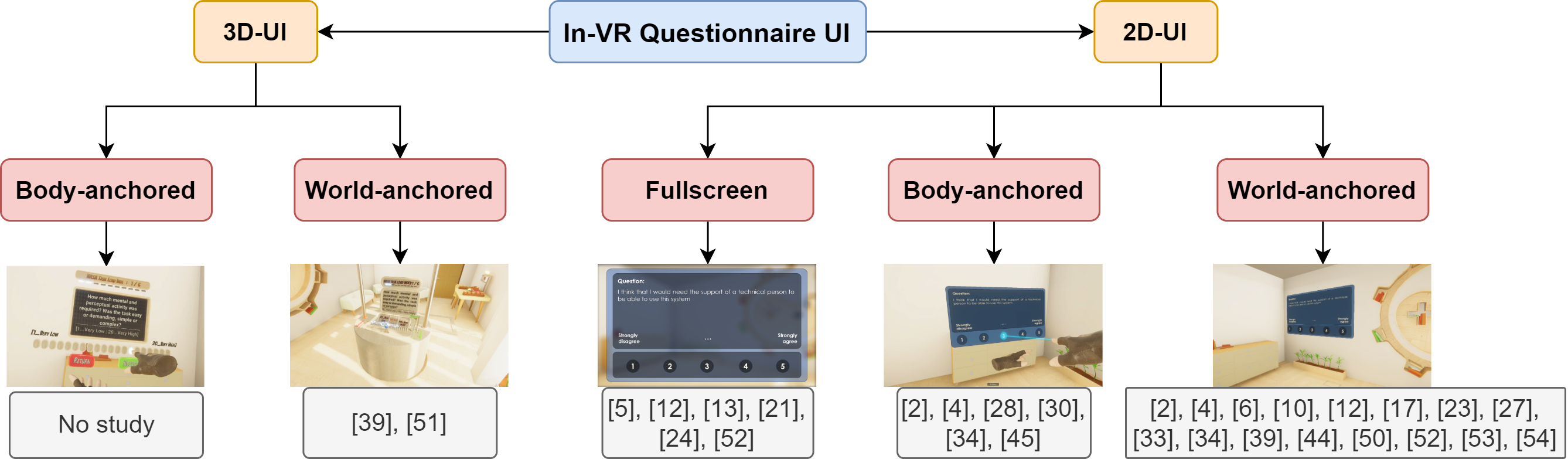}
    \caption{The taxonomy of in-VR questionnaire UI}
    \label{fig:UI_taxonomy}
\end{figure*}

Based on the reviewed literature, we can categorize the UI for implemented questionnaires in VR into two general categories: \textbf{2D UI} and \textbf{3D UI} (see \autoref{fig:UI_taxonomy}). This distinction is related to the diegetic design of UI (i.e., 2D-UI can be counted as an interface in non-diegetic space and 3D-UI can be defined as a diegetic UI because it appears in the game's context). According to these two types, developers can implement the in-VR questionnaire with different positioning approaches: full-screen, world-anchored, and body-anchored.

The full-screen approach is similar to the traditional desktop experience (i.e., it pushes the game aside and forces the user to see and interact with the questionnaire). This presentation can reduce the user's sense of presence by disconnecting them from the virtual context. The world-anchored representation can be defined as a UI element in the virtual environment that is tied to a part of the scene. It can be fully fixed (the user must align to read the content) or rotate around an edge of the UI to align with the user's view; an example of this positioning approach can be found in a study by Rzayev et al. \cite{rzayev2021reading,rzayev2018reading}. Body anchor positioning connects the questionnaire UI to a part of the virtual character (mainly the hand). The advantages of this approach are the availability of the questionnaire in all environments and the possibility to change the position of the UI on the go.

Alexandrovsky et al. \cite{alexandrovsky2020examining} used this approach to find suitable design options for in-VR questionnaires. There is also a side approach that is mainly a combination of world anchor and body anchor to take advantage of both cases. In this case, the users can trigger the user interface of the questionnaire anywhere in the virtual environment according to their preferences, but the questionnaire functions as an object anchored in the world. An example of this hybrid positioning option is described by Safikhani et al. \cite{SafikhaniIUIV} for an in-VR inventory and menu system.

We can summarize the benefits of each UI representation as the following:

\textbf{2D-UI | Fullscreen$\rightarrow$}  \textit{Benefits}: visibility of the text, focus on the questionnaire, familiar interface (similar to the traditional desktop UI), \textit{Drawbacks}: may reduce the sense of presence, detach the user from virtual experience.

\textbf{2D-UI | Body-anchored$\rightarrow$}  \textit{Benefits}: freedom in positioning and spawning, familiar interface (similar to the tradition desktop UI), \textit{Drawbacks}: more physically demanding, instability of hand during interaction, distinguishable from the rest of the scene (i.e. inconsistent with other virtual elements in the experience)

\textbf{2D-UI | World-anchored$\rightarrow$}  \textit{Benefits}: easy to interact, easy to read, familiar interface (similar to the tradition desktop UI), \textit{Drawbacks}: limited to the specific area of the scene, hard to read from distance, distinguishable from the rest of the scene)

\textbf{3D-UI | Body-anchored$\rightarrow$}  \textit{Benefits}: freedom in positioning and spawning, consistency with other virtual elements in the environment, \textit{Drawbacks}: more physically demanding, instability of hand during interaction)

\textbf{3D-UI | World-anchored$\rightarrow$}  \textit{Benefits}: easy to interact, easy to read, consistency with other virtual elements in the environment, \textit{Drawbacks}: limited to the specific area of the scene, hard to read from distance, more physical demanding than 2D-UI due to the interaction type)

The type of interaction in VR is an inseparable part of the virtual experience and is related to the type of interface. Over the years, several studies have been conducted to develop and improve interaction systems in VR. Interaction type can affect the entire user experience and change the way the user perceives this immersive environment. 

Hepperle et al. \cite{HEPPERLE2019321} investigated the impact of interaction type on user preferences and performance. Their quantitative user study considered the following three main types of user input: 2D, 3D, and speech. In their study, the 2D interface is represented by displaying 2D icons to select colors or move objects, the 3D interface is a direct interaction with virtual objects (inserting a paintbrush into a paint bucket or moving an object by grabbing it), and the speech interface can perform prior actions via voice commands. The results show that the type of interaction can affect the user experience. A 3D UI and corresponding interaction are more appropriate when perceived presence is a critical issue. A 2D interface can be chosen when accurate or fast interaction is required. Voice input is easy to learn and suitable for long text input.

\begin{figure*}
    \centering
    \includegraphics[width=\textwidth]{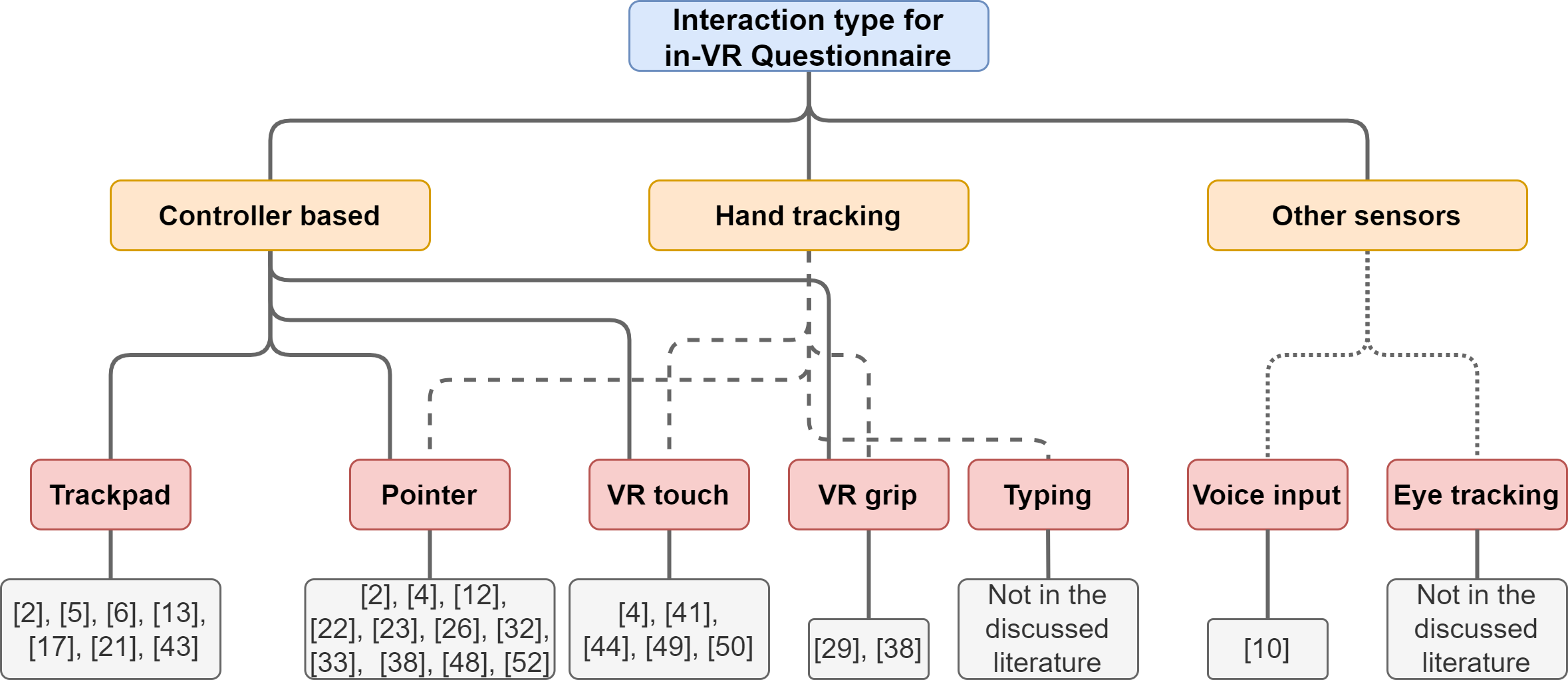}
    \caption{The interaction types with in-VR questionnaires}
    \label{fig:InteractionType}
\end{figure*}

Considering the study of Hepperle et al. \cite{HEPPERLE2019321}, the content of the VR application should be considered when deciding on UI and interactions. Accordingly, the application for in-VR questionnaires should be based on the content of the questionnaire. For example, if presence perception is a major part of the questionnaire, 3D UI and interaction can be a solution.
In \autoref{fig:InteractionType}, we categorize the possible interaction types for in-VR questionnaires. The classification is based on the input device and the resulting interaction. Some of these interaction types are not used in the studies discussed in this paper and can be a topic for future research and development. Furthermore, different input devices can be used for an identical resulting interaction (e.g., controller input and hand tracking can be used for pointer interaction), but with different usability and corresponding user experience. 
We can summarize the benefits and drawbacks of these types of interactions as the following:

\textbf{Trackpad$\rightarrow$}  \textit{Benefits}: accurate and easy to use, similar to traditional input system using gamepad or keyboard, \textit{Drawbacks}: may lead to distraction from virtual environment, difficult to manipulate when there are so many items to choose from.

\textbf{Pointer$\rightarrow$}  \textit{Benefits}: easy to implement and find template for implementation, similar to traditional input system using mouse, \textit{Drawbacks}: instability of the hands during interaction, physically demanding for long questionnaire.

\textbf{VR touch$\rightarrow$}  \textit{Benefits}: may lead to a higher perceived sense of presence, easy to interact, \textit{Drawbacks}: limited interaction range (user's virtual hand should touch the interaction area).

\textbf{VR grip$\rightarrow$}  \textit{Benefits}: may lead to a higher perceived sense of presence, consistent with the rest of virtual scene, \textit{Drawbacks}: limited interaction range (user should grab the interactive object), can be slower than others and be physically demanding.


\section{Discussion}


\textbf{RQ1:}
As shown in the Results section, there is no explicit naming convention for In-VR questionnaires. The lack of an exact keyword or abbreviation makes it difficult to research and reference this assessment tool. Accordingly, researchers cannot easily gather information about recent developments in this field. Therefore, we recommend using the term ``inVRQ'' as an abbreviation and keyword for in-VR questionnaires in future studies.


\noindent\textbf{RQ2:}
In most studies, a pointer was used to interact with the questionnaire UI.
On the other hand, based on several literature reports (e.g., Safikhani et al. \cite{safikhani2021influence} and Alexandrovsky et al. \cite{alexandrovsky2020examining}), it seems that in-VR questionnaires may have higher task requirements. Here, the question may be: Is there a relationship between interaction type and task demand? Accordingly, investigating the influence of interaction type on task demands could be an important future research topic.
Although the studies mainly used traditional 2D user interfaces, using a more consistent design approach for in-VR questionnaires may lead to a higher sense of presence, as discussed by Wagener et al. \cite{wagener2020investigating}. Considering an appropriate user interface for questionnaire presentation, as well as improving interactions in VR, may lead to increased use of in-VR questionnaires.
The majority of the presented questionnaires were limited to multiple-choice questions.
Although these types of questionnaires are typical in studies, the idea of implementing free-form questionnaires remains important.
Incorporating a free-form questionnaire can be helpful, especially after the experience when users are required to provide their personal feedback.
Discouraging the implementation of free-form questionnaires may be related to the difficulty of typing in VR. Here, it will be of interest to find a suitable solution for another type of user input.

\noindent\textbf{RQ3:}
The ability to conduct VR studies remotely may facilitate the study procedure. In addition, the increasing number of consumer HMDs will likely make remote evaluation more feasible. In this case, using the in-VR questionnaire as part of a distance learning package may be a solution to facilitate VR studies and increase the number of participants.
Based on our findings, we can recommend the following points for future studies:
(1) Investigating the influence of interaction type on task load during the filling of in-VR questionnaires 
(2) Define and implement a standard design for in-VR questionnaires, considering the consistency with the virtual environment 
(3) Study on appropriate user input for different kinds of questions, such as free-form ones.
(4) Design and development of an all-in-one toolkit for importing, interacting, and exporting in-VR questionnaires with the possibility of running remote studies

\RVone{ As an overview, in-VRQ questionnaires can improve the overall user experience and consistency of results by reducing BIPs and eliminating the interruptions caused by switching between the real and virtual worlds. They can help in conducting the study remotely. However, the development of an in-VR questionnaire is less relevant due to the implementation challenges and difficulties in finding relevant literature. 
We believe that this categorized structure for interaction types and UI design options could help developers during implementation. On the other hand, our recommendation of a general term, inVRQ, could help future studies refer to these questionnaires and lead to clearer results in future literature collections. 
}


\section{Conclusion}
In VR research, measuring parameters related to the VR experience, such as presence, can be affected by an interruption in the experience that may lead to inaccuracy in the results. According to the literature, using questionnaires embedded in VR can lead to less break in presence.
In this study, we conducted a literature review on the use of In-VR questionnaires. We reviewed a total of 25 articles from 2014-2021 in which an In-VR questionnaire was used either as an assessment tool or as a topic for study. 

This study can be extended by collecting literature from more databases.
We also did not perform a quality assessment for the content of the reviews because we were only interested in the application of the in-VR questionnaires. The use of quality assessment tools, such as JBI, may help future research to ensure the quality of reported studies.
The results of studies on the application of in-VR questionnaires in comparison to the out-VR type indicate a positive response of participants towards in-VR ones.
In this case, the consistency of in-VR questionnaire design with the virtual environment from both interaction and UI may lead to a higher perceived sense of presence.
However, comparing the task load in in-VR and out-VR questionnaires shows higher values in the case of in-VR ones.
Accordingly, studies on the influencing factors on increasing in-VR experience task load can be investigated to improve the usability and acceptability of this type.

The difficulties in the design and implementation of in-VR questionnaires can be a barrier to an increase in their application.
Considering this point, the development of comprehensive toolkits that makes the implementation of it faster and simpler can lead to more acceptance among researchers.
In addition, according to the increased number of consumer-based HMDs in recent years, developing a toolkit with the possibility of remote evolution can help studies to have more participants.
These kinds of toolkits can also be helpful in a situation where accessing users is not possible or limited, such as the current COVID-19 pandemic. In this case, an all-in-one VR questionnaire package may be used more in research communities. This taxonomy of in-VR questionnaires presented in this paper may provide a useful overview of common ways to present and interact with questionnaires in VR.

%
%
%
\bibliographystyle{splncs04}
\bibliography{01_Bib}

\end{document}